\documentclass{svmult}
 
\usepackage{amsmath}
\usepackage{amsfonts}
\usepackage{euscript}

\usepackage{amsbsy}
\usepackage{amssymb}
 
\smartqed

\newcommand{\R}{\mathbb{R}}
\newcommand{\N}{\mathbb{N}}
\newcommand{\Z}{\mathbb{Z}}

\newcommand{\B}{\mathbb{B}}

\newcommand{\lb}{\linebreak}

\newcommand{\cl}[1]{\left\lceil #1 \right\rceil}
\newcommand{\fl}[1]{\left\lfloor #1 \right\rfloor}

\newcommand{\ewa}{e^{\mathrm{wor-avg}}}

\newcommand{\ea}{e^{\mathrm{avg}}}
\newcommand{\oSM}{S_{M,1}}

\newcommand{\oSMq}{S_{M,q}}
\newcommand{\oSMqp}{S^{\,\prime}_{M,q}}

\newcommand{\osa}{\overline{s}_a}
\newcommand{\usa}{\underline{s}_a}

\begin{document}

\title{ Quantum Boolean Summation with Repetitions in the
  Worst-Average Setting \thanks{The research of the second and third
    coauthors was supported in part by the National Science
    Foundation (NSF) and by the Defense Advanced Research Agency
    (DARPA) and Air Force Research Laboratory under agreement
    F30602-01-2-0523.}}

\author{
Stefan Heinrich \inst{1} \and Marek Kwas \inst{2,3}  \and
Henryk Wo\'zniakowski \inst{2,3}
}

\institute{
Universit\"a{}t Kaiserslautern, FB Informatik, Postfach 3049,
D-67653,\\
Kaiserslautern, Germany, \and
Department of Computer Science, Columbia University,\\ 
New York, NY 10027, USA, \and 
Institute of Applied Mathematics and Mechanics, University of Warsaw,\\
ul. Banacha 2, 02-097 Warszawa, Poland, \\
emails: \emph{ heinrich@informatik.uni-kl.de, \{marek,
  henryk\}@cs.columbia.edu}}

\titlerunning{Quantum Boolean Summation  in the
  Worst-Average Setting}

\maketitle

\begin{abstract}
  We study the quantum summation (\textbf{QS}) algorithm of Brassard,
  H\o{}yer, Mosca and Tapp, 
  see \cite{BHMT}, which approximates the
  arithmetic mean of a Boolean function defined on $N$ elements. 
  We present sharp error bounds of the \textbf{QS}
  algorithm in the worst-average setting with the 
  average performance measured in the $L_q$ norm, $q \in [1,\infty]$.
  
  We prove that the \textbf{QS} algorithm with $M$ quantum queries,
  $M<N$, has the worst-average error bounds of the form $\Theta(\ln
  M/M)$ for $q=1$, $\Theta(M^{-1/q})$ for $q\in (1,\infty)$, and
  is equal to $1$ for $q=\infty$. We also discuss the asymptotic constants
  of these estimates.
  
  We improve the error bounds by using the \textbf{QS} algorithm with
  repetitions. Using the number of repetitions which is independent of
  $M$ and linearly dependent on~$q$, we get the error bound of order
  $M^{-1}$ for any $q \in [1,\infty)$. Since $\Omega(M^{-1})$ is a
  lower bound on the worst-average error of any quantum algorithm with
  $M$ queries, the \textbf{QS} algorithm with repetitions is optimal
  in the worst-average setting.
\end{abstract}

\section{Introduction}
\label{sec:intro}

The quantum summation (\textbf{QS}) algorithm of Brassard, H\o{}yer,
Mosca and Tapp computes an approximation to the arithmetic mean of the
values of a Boolean function defined on a set of $N=2^n$ elements. An
overview of the \textbf{QS} algorithm and its basic properties is
presented in the first two sections of~\cite{KW}.  In Section
\ref{sec:QS-algo} we remind the reader of the  facts concerning the
\textbf{QS} algorithm that are needed in this paper.

The \textbf{QS} algorithm enjoys many optimality properties and has
many applications.  It is used for the summation of real numbers which
in turn is an essential part for many continuous problems such as
multivariate and path integration, and multivariate approximation. The
knowledge of the complexity of the quantum summation problem allows us
to determine the quantum complexity of many continuous problems, such
as those mentioned above, see~\cite{Novak} and a recent survey
\cite{Stefannew}.

The \textbf{QS} algorithm has been studied in the two error settings so far:
\begin{itemize}
\item worst-probabilistic in \cite{BHMT,KW},
\item average-probabilistic in \cite{KW}.
\end{itemize}
These settings are defined by taking the worst case/average performance with
respect to all Boolean functions and the probabilistic performance with respect
to outcomes of the \textbf{QS} algorithm. 

It turns out that the \textbf{QS} algorithm is optimal in these two
settings.  The corresponding lower bounds for the Boolean summation
problem were shown in \cite{NayakWu} for the worst-probabilistic
setting, and in \cite{Papa} for the average-probabilistic setting.  In
particular, we know that the \textbf{QS} algorithm with $M$ quantum
queries, $M<N$,  has the error bound of order $M^{-1}$ in the
worst-probabilistic setting.

In this paper we study the \emph{ worst-average} setting.
In this setting, we take the worst case performance over all Boolean 
functions and the average performance over all outcomes of the 
\textbf{QS} algorithm. The average performance is measured in 
the $L_q$ norm, $q \in [1,\infty]$. This setting is 
analogous to the randomized (Monte Carlo) setting used for 
algorithms on a classical computer.
The worst-average setting also seems to be quite natural 
for the analysis of quantum algorithms. 

As we shall see, the results depend on the choice of $q$. Obviously,
for larger~$q$, the effect of the average behavior becomes less
significant.  In fact, the limiting case, $q=\infty$, leads to the
deterministic case (modulo sets of measure zero). Not surprisingly,
for $q=\infty$, the results are negative.

In what follows we indicate error bounds for large $M$. Since we
always assume that $M<N$, this means that for $M$ tending to infinity
we also let $N$ tend to infinity. To make error bounds independent of
$N$, we take the supremum over $N>M$ in the corresponding definitions
of the errors. When we speak about the sharpness of error bounds, we
usually take a large $M$ and select a still larger $N$ and a Boolean
function for which the presented error bound is sharp.

The worst-average error $\ewa_q(M)$ of the \textbf{QS} algorithm with
$M$ quantum queries satisfies:
\begin{itemize}
\item For $q=1$, we have $\displaystyle{ \ewa_1(M)=
    \Theta\bigg(\frac{\ln M}M\bigg)}$.  Furthermore, the asymptotic
  constant is $2/\pi$ for $M-2$ divisible by $4$.
    
\item For $q\in (1,\infty)$, we have $\displaystyle{ \ewa_q(M) =
    \Theta\bigg(\frac1{M^{1/q}}\bigg)}$.  Furthermore, the
  asymptotic constant is approximately $\big(\int_0^{\pi}
  \sin^{q-2}(x)dx/\pi\big)^{1/q}$ for $M-2$ divisible by $4$ and $q$ close to $1$.
\item For $q=\infty$,  we have $\ewa_\infty(M)=1$.
\end{itemize}

The error bounds of the \textbf{QS} algorithm are improved by the use
of repetitions. Namely, we repeat the \textbf{QS} algorithm $2n +1$
times and take the median of the outputs obtained as the final output.
This procedure boosts the success probability of the approximation at
the expense of the number of quantum queries. We show that with $n$
independent of $M$ and linearly dependent on~$q$, we decrease the
\textbf{QS} algorithm error to be of order $M^{-1}$. Hence, the use of
repetitions is particularly essential for large $q$ since we change
the error bound $O(M^{-1/q})$ without repetitions to the error bound
$O(M^{-1})$ with repetitions. The constant in the last big $O$
notation is absolute and does not depend on $q$ and $M$.

The error bound of order $M^{-1}$ is optimal.
This follows from the use of, for instance, Chebyshev's inequality
and the fact that the lower bound $\Omega(M^{-1})$ is sharp in the
worst-probabilistic setting, see also \cite{Papa}. 
Hence, the \textbf{QS} algorithm with repetitions is optimal
in the worst-average setting. 

\section{Quantum Summation Algorithm}
\label{sec:QS-algo}

The quantum summation \textbf{QS} algorithm of Brassard, H\o{}yer, 
Mosca and Tapp, see \cite{BHMT}, approximates the mean 
 \begin{equation*}
   a_f=\frac1N\sum_{i=0}^{N-1} f(i)
 \end{equation*}
of a Boolean function $f:\{0,1, \ldots, N-1\} \rightarrow \{0,1\}$. 
Without loss of generality we assume that $N$ is a power of two. 

The \textbf{QS} algorithm uses $M-1$ quantum queries. The only
interesting case is when $M$ is much smaller than $N$. The
\textbf{QS} algorithm returns an index $j\in\{0,1,\dots,M-1\}$ with
probability
\begin{equation*}
p_f(j)= \frac{\sin^2( M
  \theta_{a_f})}{2M^2}\;\bigg(\sin^{-2}\bigg(\frac{\pi(j-\sigma_{a_f})}M\bigg)
  + \sin^{-2}\bigg(\frac{\pi(j+\sigma_{a_f})}M\bigg)\bigg) ,    
\end{equation*}
see \cite{KW} for the detailed analysis of the \textbf{QS} algorithm.
Here
\begin{align*}
  \theta_{a_f}=\arcsin \sqrt{a_f} \qquad \text{and} \qquad \sigma_{a_f} =
  \frac{M}{\pi}\, \theta_{a_f}.
\end{align*}
We will also be using 
\begin{equation*}
  s_{a_f} =\min \big\{ \cl{\sigma_{a_f}} - \sigma_{a_f}, \sigma_{a_f} -
  \fl{\sigma_{a_f}} \big\}.
\end{equation*}
Clearly, $s_{a_f}\in [0,\tfrac12]$ and $s_{a_f}=0$ iff $\sigma_{a_f}$ is an
integer. We shall usually drop the subscript $f$ and denote
$\theta_a=\theta_{a_f}$, $\sigma_a=\sigma_{a_f}$, $s_a=s_{a_f}$ when
$f$ is clear from the context.  

Knowing the index $j$, we compute the output
\begin{equation*}
\bar a_f(j)\,=\,\sin^2\left(\frac{\pi j}M\right)
\end{equation*}
on a classical computer. The error is then given by
\begin{equation}
  \label{eq:prob}
|a_f - \bar a_f(j)|=\left|\sin\bigg(\frac{\pi(j-\sigma_{a_f})}M\bigg)
                          \sin\bigg(\frac{\pi(j+\sigma_{a_f})}M\bigg)
                          \right|. 
\end{equation}

As in \cite{KW}, we let $\mu(\cdot,f)$ denote the measure on the set of all
possible outcomes of the \textbf{QS} algorithm which is defined as
\begin{equation*}
 \mu(A,f)=\sum_{j \in A} p_f(j) \qquad \forall A \subset \{0,1,\ldots, 
M-1\}.  
\end{equation*}
Let $\mathcal{A}_M$ denote the set of all possible outputs of the
\textbf{QS} algorithm with $M-1$ queries, i.e.,
\begin{equation*}
  \mathcal{A}_M=\bigg\{ \sin^2\bigg(\frac{\pi j}M\bigg):\; j=0,1,\ldots,
  M-1\bigg\}. 
\end{equation*}
Let
\begin{equation*}
  \rho_f(\alpha) = \mu \bigg( \bigg\{ j \in \{ 0,1,\ldots, M-1\} :
  \sin^2\bigg(\frac{\pi j}M \bigg)=\alpha \bigg\}, f \bigg) \quad \forall
  \alpha \in 
  \mathcal{A}_M,
\end{equation*}
denote the probability of the output $\alpha$.  Note that
$\alpha=\sin^2(\pi j /M) \lb = \sin^2 (\pi (M-j)/M)$. Hence if $j \ne 0$
and $j \ne M/2$ then $\rho_f(\alpha) = p_f(j) + p_f(M-j)$.

In what follows  we let $\B_N$ denote the set of all Boolean functions 
defined on $\{0,1,\dots,N-1\}$.

\section{Performance Analysis}
\label{sec:perform} 
The error of the \textbf{QS} algorithm in the worst-probabilistic and
average-probab- ilistic settings has been analyzed in \cite{BHMT,KW}. 
In this paper we analyze the error of the \textbf{QS} algorithm in
the  worst-average setting. This corresponds to the worst case
performance with respect to all Boolean functions from $\B_N$ and the
average performance 
with respect to all outcomes. This average performance is measured
by the expectation in the $L_q$ norm, $q\in [1,\infty]$,
with respect to the probability measure of all outcomes provided by
the \textbf{QS} algorithm.  As mentioned before, we make the
worst-average error independent of $N$ by taking the supremum over
$N>M$. That is,  the worst-average error 
is defined as:
\begin{itemize}
\item  for $q\in[1,\infty)$,
\begin{equation*}
  \ewa_q(M)= \sup_{N>M}\,\max_{f \in \B_N}\left(\sum_{j=0}^{M-1} p_f(j)\,|a_f-
    \bar a_f(j)|^q\right)^{1/q},
\end{equation*}
\item  for $q=\infty$,
\begin{equation*}
  \ewa_{\infty}(M)= \sup_{N>M}\,\max_{f \in \B_N}
\max_{j:\,p_f(j)>0}|a_f-\bar a_f(j)|.
\end{equation*}
\end{itemize}

It is easy to check that for $q=\infty$, the \textbf{QS} algorithm
behaves badly.  Indeed, if $M$ is odd, we can take $f$ with all values
one, and then $a_f=1$, $p_f(0)=1/M^2$ and $\bar a_f(0)=0$. Hence
$\ewa_{\infty}(M)=1$.  If $M$ is even, we take $f$ with only one value
equal to $1$, and then $a_f=1/N$, $p_f(M/2)>0$ and $\bar a_f(M/2)=1$.
Hence, $|a_f-\bar a_f(M/2)|=1-1/N$ and $\ewa_{\infty}(M)=1$. 

That is why in the rest of the paper we consider $q\in [1,\infty)$. 
As we shall see the cases $q>1$ and $q=1$ will require a
different analysis and lead to quite different results. 

\subsection{Local Average Error}
\label{sec:LAErr}

We  analyze the local average error for a fixed function $f \in
\B_N$ for $1 \le q < \infty$, 
\begin{equation}
  \label{eq:e-avg}
  \ea_q(f,M)=\bigg( \sum_{j=0}^{M-1} p_f(j)\,|a_f- \bar a_f(j)|^q
  \bigg)^{1/q}= \bigg( \sum_{\alpha \in \mathcal{A}_M} \rho_f (\alpha)
  | a_f - \alpha|^q\bigg)^{1/q}.
\end{equation}

We first analyze the case  $q>1$. 
\begin{theorem}
\label{thm:ErrAvgLoc_q>1}
Let $q\in(1,\infty)$. Denote $a=a_f$. If $\sigma_a \in \Z$ then
$\ea_q(f,M)=0$. If $\sigma_a \notin \Z$ then
\begin{multline}
%  \label{eq:ErrAvgLoc_q>1}
  \Bigg| \ea_q(f,M)^q - \frac{\sin^2(\pi s_a)}{M\pi} \int_{\pi \osa /
    M}^{\pi - \pi \usa / M} \sin(x)^{q-2} \big|\sin(x + 2
  \theta_a)\big|^q \,dx \Bigg| \le \\ (1+2(1-\delta_{q,2})) \frac{\pi^{q-1}
    \sin(\pi s_a)}{M^{q}} + \frac{\sin^2 (\pi
    s_a)}{M^2}\Bigg(2(1-\delta_{q,2}) + q \int_0^\pi \sin^{q-2}(x)\,dx
  \Bigg),
\end{multline} 
with $\usa=\fl{\sigma_a} -\sigma_a$ and $\osa=
\sigma_a-\cl{\sigma_a}$.\footnote{Note that the
  last integral is finite. It is obvious for $q\ge2$.  For
  $q\in(1,2)$, the only singularities are at the boundary points and
  are of the form $x^{q-2}$ for $x$ approaching $0$. The function
  $x^{q-2}$ is integrable since $q>1$.}

% \begin{multline}

%   \ea_q(f,M) = \frac1{M^{1/q}} \bigg[\frac{\sin^2(\pi s_a)}{2\pi} \bigg(
%   \int_0^\pi 
%   \sin^{q-2}(x) \big|\sin(x+2 \theta_a) \big|^q\,dx \\
%   + \int_0^\pi \sin^{q-2}(x) \big|\sin(x-2 \theta_a)
%   \big|^q\,dx\bigg)\bigg]^{1/q}\big(1+o(1)\big),
% \end{multline}
% with $s_a\in (0,\tfrac12]$, and $o(1)$ denotes a term tending to $0$ for
% large $M$.
\end{theorem}
\begin{proof}
  If $\sigma_a \in \Z$ then it is shown in \cite{KW} that there exists
$\alpha \in \mathcal{A}_M$ such that $\alpha=a_f$ and
$\rho_f(\alpha)=\delta_{\alpha,a_f}$ for all $\alpha \in
\mathcal{A}_M$. Then $\ea_q(f,M)=0$ as claimed.  

Assume that $\sigma_a \notin \Z$. Using the form
of $p_f(j)$ from Section~\ref{sec:QS-algo}, we rewrite~(\ref{eq:e-avg}) as
\begin{multline*}
  \big(\ea_q(f,M)\big)^{q} = \sum_{j=0}^{M-1}
  \frac{\sin^2(M\theta_a)}{2M^2} \bigg( \bigg|
  \sin\bigg(\frac{\pi(j-\sigma_a)}M\bigg)\bigg|^{q-2} 
   \bigg|\sin\bigg(\frac{\pi(j+\sigma_a)}M\bigg)\bigg|^q \\
  + \bigg| \sin\bigg(\frac{\pi(j+\sigma_a)}M\bigg)\bigg|^{q-2}
  \bigg|\sin\bigg(\frac{\pi(j-\sigma_a)}M\bigg) \bigg|^q \bigg).
\end{multline*}
We have
\begin{multline*}
  \sum_{j=0}^{M-1}
   \bigg| \sin\bigg(\frac{\pi(j+\sigma_a)}M\bigg)\bigg|^{q-2}
  \bigg|\sin\bigg(\frac{\pi(j-\sigma_a)}M\bigg) \bigg|^q \\
  =\sum_{j=1}^{M}
   \bigg| \sin\bigg(\frac{\pi(M-j+\sigma_a)}M\bigg)\bigg|^{q-2}
  \bigg|\sin\bigg(\frac{\pi(M-j-\sigma_a)}M\bigg) \bigg|^q.
\end{multline*}
Using the $\pi$-periodicity of $|\sin x|$, we see that the last sum is
equal to 
\begin{multline*}
  \sum_{j=1}^{M}
   \bigg| \sin\bigg(\frac{\pi(j-\sigma_a)}M\bigg)\bigg|^{q-2}
  \bigg|\sin\bigg(\frac{\pi(j+\sigma_a)}M\bigg) \bigg|^q \\
 =\sum_{j=0}^{M-1}
   \bigg| \sin\bigg(\frac{\pi(j-\sigma_a)}M\bigg)\bigg|^{q-2}
  \bigg|\sin\bigg(\frac{\pi(j+\sigma_a)}M\bigg) \bigg|^q. 
\end{multline*}
Therefore
\begin{equation}
\label{eq:oSMq}
\ea_q(f,M)^{q} =  \frac{\sin^2(M\theta_a)}{M^2} \oSMq
\end{equation}
with
\begin{multline*}
  \oSMq=\sum_{j=0}^{M-1} \bigg|
  \sin\bigg(\frac{\pi(j-\sigma_a)}M\bigg)\bigg|^{q-2}
  \bigg|\sin\bigg(\frac{\pi(j+\sigma_a)}M\bigg) \bigg|^q \\
  = \sum_{j=0}^{M-1} \bigg|
  \sin\bigg(\frac{\pi j}M - \theta_a\bigg)\bigg|^{q-2}
  \bigg|\sin\bigg(\frac{\pi j}M +\theta_a\bigg) \bigg|^q.
\end{multline*}
We split $\oSMq$ as
\begin{equation*}
\label{eq:osmq_prime}
  \oSMqp=\oSMq- \bigg| \sin\bigg(\frac{\pi \fl{\sigma_a}}M -
  \theta_a\bigg)\bigg|^{q-2} \bigg|\sin\bigg(\frac{\pi \fl{\sigma_a}}M
  +\theta_a\bigg) \bigg|^q.
\end{equation*}
Observe  that $\frac\pi{M} \oSMqp$ is the rectangle formula for 
approximating the integral
\begin{equation*}
 \int_{[0,\pi] \setminus [\pi \fl{\sigma_a} /M,\pi \cl{\sigma_a} /M]}
 \big|\sin(x - \theta_a)\big|^{q-2} \big|\sin(x + \theta_a)\big|^q\, dx.
\end{equation*}

 The error of the rectangle quadrature for $k \in \N$ and an
absolutely continuous function $f: [a,b] \rightarrow \R$ whose first
derivative belongs to $L_1([a,b])$  satisfies
\begin{equation}\label{rectangle}
  \bigg| \int_a^b f(x)\, dx - \frac{b-a}{k}\sum_{j=0}^{k-1} f\bigg(a+
  j\frac{b-a}k\bigg) \bigg| \le \frac{b-a}k \int_a^b
  \big|f\,'(x)\big|\, dx.    
\end{equation}
Thus defining ${h}(x) = \big|\sin(x - \theta_a)\big|^{q-2}
\big|\sin(x + \theta_a)\big|^q$ and $D_a =[0,\pi] \setminus [\pi
\fl{\sigma_a} /M,\pi \cl{\sigma_a} /M]$ and using the error formula above
for the subintervals $[0, \pi\fl{\sigma_a}/M)$ and $(\pi\cl{\sigma_a}/M,
\pi]$, we get
\begin{equation*}
   \bigg|\frac\pi{M}\oSMqp - \int_{D_a} {h}(x) \, dx\bigg| \le
   \frac\pi{M} \int_{D_a} \big|{h}\,'(x)\big| \, dx.
\end{equation*}
Define ${H}(x) = {h}(x+\theta_a)=
\big|\sin(x)\big|^{q-2} \big|\sin(x + 2 \theta_a)\big|^q$ and
$\Delta_a = [-\theta_a,\pi-\theta_a] \setminus [\pi (\fl{\sigma_a}
-\sigma_a) /M,\pi (\sigma_a-\cl{\sigma_a}) /M]$ . We have
\begin{equation*}
  \int_{D_a} {h}(x) \, dx =  \int_{\Delta_a}{H}(x)
  \, dx, \qquad 
  \int_{D_a} \left|{h}\,'(x)\right| \, dx  = \int_{\Delta_a}
  \left|{H}\,'(x) \right|\, dx.
\end{equation*}
and by the $\pi$-periodicity of the integrand
${H}$ we have
\begin{multline*}
  \int_{\Delta_a} {H}(x) \, dx = \int_{-\theta_a}^{- \pi \usa
    /M} {H}(x) \, dx + \int_{\pi \osa /M}^{\pi - \theta_a}
  {H}(x) \, dx \\ = \int_{\pi - \theta_a}^{\pi - \pi \usa /M}
  {H}(x) \, dx + \int_{\pi \osa /M}^{\pi - \theta_a}
  {H}(x) \, dx = \int_{\pi \osa /M}^{\pi - \pi \usa/M}
  {H}(x) \, dx. 
\end{multline*}
 Analogously, 
\begin{equation*}
\int_{\Delta_a} \left|{H}\,'(x) \right|\, dx  =
  \int_{\pi \osa /M}^{\pi - \pi \usa /M} |{H}\,'(x) | \, dx. 
\end{equation*}
 For $x \in [\pi \osa / M, \pi - \pi \usa
/M]$ the sine is positive and 
\begin{equation*}
  \left|{H}\,'(x) \right| \le |q-2| \sin^{q-3}(x) |\cos (x)|
  + q \sin^{q-2}(x). 
\end{equation*}
It is easy to check that for $q \ne 2$ we have 
\begin{multline*}
  \int_{\pi \osa/M}^{\pi -
    \pi \usa/M} |q-2| \sin^{q-3}(x) |\cos(x)| \, dx \\ = |q-2| \bigg(
    \int_{\pi \osa/M}^{\pi/2} \sin^{q-3} (x) \, d \sin(x) - 
    \int_{\pi/2}^{\pi - \pi \usa/M}   \sin^{q-3} (x) \, d
    \sin(x)\bigg)\\
 = \frac{|q-2|}{q-2} \Bigg( 2 - \sin^{q-2}\bigg(\frac{\pi \osa}{M}\bigg) - 
   \sin^{q-2}\bigg(\pi  - \frac{\pi \usa}{M}\bigg) \Bigg). 
\end{multline*}
From this we get
\begin{multline*}
  \int_{\pi \osa/M}^{\pi - \pi \usa/M} \left|H\,'(x) \right| \, dx \le
  (1-\delta_{q,2})\Bigg( 2 + \sin^{q-2} \bigg(\frac{\pi\osa}M\bigg) \\ 
  + \sin^{q-2} \bigg(\frac{\pi \usa }M\bigg) \Bigg) + q \int_0^\pi
  \sin^{q-2}(x)\,dx.
\end{multline*}
We then finally get 
\begin{multline*}
  \Bigg| \frac\pi{M} \oSMq - \int_{ \pi \usa /M}^{\pi - \pi \osa/M}
  {H}(x) \, dx \Bigg|  
 \le \frac\pi{M} \Bigg(
  (1-\delta_{q,2})\bigg( 2 + \sin^{q-2} \bigg(\frac{\pi\osa}M\bigg) 
  \\ + \sin^{q-2} \bigg(\frac{\pi \usa }M\bigg) \bigg)  + \sin^{q-2}
  \bigg(\frac{\pi \usa }M\bigg)+ q \int_0^\pi \sin^{q-2}(x)\,dx \Bigg).
\end{multline*}
Observe also that 
\begin{equation*}
  \sin (\pi s_a) =  \sin(\pi \usa)=  \sin(\pi\osa ).
\end{equation*}
Since $\sin(x)/[M\sin(x/M)] \le 1$ for  $x
\in (0, \pi]$, we get
\begin{multline*}
  \Bigg| \frac{\pi \sin(\pi s_a)}{M} \oSMq - \sin(\pi s_a)
  \int_{\pi \osa / M}^{\pi -
    \pi\usa / M} {H}(x) \, dx \Bigg|
  \\ \le \pi (1-\delta_{q,2}) \Bigg( \sin^{q-1} \bigg(\frac{\pi
  \osa}M\bigg) + \sin^{q-1} \bigg(\frac{\pi \usa }M\bigg)\Bigg) + \pi
  \sin^{q-1} \bigg(\frac{\pi \usa }M\bigg) \\ + \frac{\pi\sin (\pi
    s_a)}{M}\Bigg(2(1-\delta_{q,2}) + q \int_0^\pi \sin^{q-2}(x)\,dx
  \Bigg).
\end{multline*}
Using $\sin(\pi\osa/M)\le\pi/M$ we obtain 
\begin{multline*}
  \Bigg| \frac{\pi \sin(\pi s_a)}{M} \oSMq - \sin(\pi s_a)
  \int_{\pi \osa /M}^{\pi -
    \pi \usa / M} {H}(x)  \, dx \Bigg|
  \\ \le (1+2(1-\delta_{q,2})) \frac{\pi^q}{M^{q-1}} + \frac{\pi\sin (\pi
    s_a)}{M}\Bigg(2(1-\delta_{q,2}) + q \int_0^\pi \sin^{q-2}(x)\,dx
  \Bigg).
\end{multline*}

Finally, since  $\sin^2(M\theta_a)=\sin^2(\pi s_a)$, we complete the proof
by using the estimate of $\oSMq$ in  (\ref{eq:oSMq}). \qed

% We can then write 
% \begin{equation*}
%   \frac1M \oSMq = \frac{1+o(1)}\pi  \int_0^\pi  \sin^{q-2}(x)
%   \big|\sin(x+2\theta_a)\big|^q\,dx
% \end{equation*}
% for large~$M$.  Analogously we get
% \begin{multline*}
%   \frac1{M}\uSMq:=\frac1{M}\sum_{j=0}^{M-1} \bigg|
%   \sin\bigg(\frac{j+\sigma_a}M\bigg) \bigg|^{q-2}
%   \bigg|\sin\bigg(\frac{j-\sigma_a}M\bigg) \bigg|^q \\
%   = \frac1{M} \sum_{j=0}^{M-1} \bigg|
%   \sin\bigg(\frac{j+\sigma_a}M\bigg) \bigg|^{q-2}
%   \bigg|\sin\bigg(\frac{j+\sigma_a}M -2\theta_a\bigg) \bigg|^q \\
% = \frac{1+o(1)}\pi \int_0^\pi  \sin^{q-2}(x) 
%   \big|\sin(x-2\theta_a)\big|^q\,dx. 
% \end{multline*}
% Combining the estimates and the fact that $\sin(M \theta_a)= \sin (\pi
% s_a)$ we get (\ref{eq:ErrAvgLoc_q>1}).
\end{proof}
Theorem \ref{thm:ErrAvgLoc_q>1} implies the following corollary.

\begin{corollary}
\label{cor:ErrAvgLoc_q>1}
Let $q\in(1,\infty)$. If $\sigma_a \in \Z$ then
$\ea_q(f,M)=0$. If $\sigma_a \notin \Z$ then
\begin{multline}
  \label{eq:ErrAvgLoc_q>1}
  \ea_q(f,M) = \frac1{M^{1/q}} \Bigg[\frac{\sin^2(\pi s_a)}{\pi}
  \Bigg( \int_0^\pi
  \sin^{q-2}(x) \big|\sin(x+2 \theta_a) \big|^q\,dx \\
  + O\left(\frac{\sin(\pi
      s_a)}{M^{\min(1,q-1)}}\right)\Bigg)\Bigg]^{1/q},
\end{multline}
with $s_a\in (0,\tfrac12]$, and the factor in the big $O$ notation is
independent of $f$ from~$\B_N$, and also independent of $N$. 
\end{corollary}

%\vspace{12pt}

We now consider the case $q=1$ and  present estimates of
$\ea_1(f, M)$ in the following lemma.
\begin{lemma}
\label{lem:ErrAvg}
Let $a=a_f$. If  $\sigma_a \in \Z$ then $\ea_1(f,M)=0$. If $\sigma_a
\notin \Z$ then  
\begin{equation}
  \label{eq:ErrAvg}
   \bigg|\ea_1(f,M) - \frac{\sin^2 (\pi s_a) \sin(2 \theta_a)}{M} 
   \,\Sigma_{M,a} \bigg| \le \frac{\sin^2 ( \pi s_a)}{M} |\cos (2 \theta_a)|,
\end{equation}
where $s_a \in (0,\frac12]$, and 
\begin{equation*}
  \Sigma_{M,a}= \frac{1}{M} \sum_{j=0}^{M-1} \bigg|
  \cot\bigg(\frac{\pi(j+s_a)}M\bigg)\bigg|.
\end{equation*}
\end{lemma}

\begin{proof}
  The case $\sigma_a \in \Z$ can be proved as in Theorem
  \ref{thm:ErrAvgLoc_q>1}.  Assume that $\sigma_a \notin \Z$. Using the form of
  $p_f(j)$ from Section \ref{sec:QS-algo}, we have
\begin{equation*}
\ea_1(f, M) =\sum_{j=0}^{M-1} \frac{\sin^2( M
  \theta_a)}{2M^2}\left( \left| \frac{\sin(\pi(j+\sigma_a)/M)}{
  \sin(\pi(j-\sigma_a)/M)} \right|+  \left| \frac{\sin(\pi(j-\sigma_a)/M)}{
  \sin(\pi(j+\sigma_a)/M)} \right| \right)
\end{equation*}
As in the proof of Theorem \ref{thm:ErrAvgLoc_q>1} we conclude that
\begin{equation*}
   \ea_1(f,M) = \frac{\sin^2 (M \theta_a)}{M^2}  \oSM, 
\end{equation*}
where
\begin{equation*}
  \oSM =\sum_{j=0}^{M-1} \left| \frac{\sin(\pi(j+\sigma_a)/M)}{
  \sin(\pi(j-\sigma_a)/M)} \right|=\sum_{j=0}^{M-1} \left|
  \frac{\sin(\pi(j-\cl{\sigma_a} + \osa)/M+ 2\theta_a )}{
  \sin(\pi(j-\cl{\sigma_a}+\osa)/M)} \right|,
\end{equation*}
with $\osa = \cl{\sigma_a} - \sigma_a$.  Changing the index~$j$ in the
second sum to $j-\cl{\sigma_a}$, and using periodicity of the sine, we
get
\begin{equation*}
  \oSM =\sum_{j=0}^{M-1} \left| \frac{\sin(\pi(j+\osa)/M+ 2 \theta_a)}{
  \sin(\pi(j+\osa)/M)} \right|
\end{equation*}
and consequently
\begin{equation*}
  \oSM =\sum_{j=0}^{M-1} \left|\, \cos(2\theta_a) +
    \sin(2\theta_a) \cot\bigg(\frac{\pi(j+\osa)}M\bigg)\, \right|.
\end{equation*}
Using the triangle inequality twice, we obtain
\begin{equation*}
\left|\,\oSM-  \sin(2\theta_a) \sum_{j=0}^{M-1} \left|
\,   \cot\bigg(\frac{\pi(j+\osa)}M\bigg) \, \right|\right| \le
M|\cos(2\theta_a)|. 
\end{equation*}
Let $\usa = \sigma_a - \fl{\sigma_a}$. Observe that $\usa = 1 - \osa$.
Since the cotangent is $\pi$-periodic and the function
$|\cot(\pi(\cdot)/M)|$ is even, we get
\begin{equation*}
 \sum_{j=0}^{M-1} \left| \cot\bigg(\frac{\pi(j+\osa)}M\bigg) \right|=
  \sum_{j=0}^{M-1} 
  \left| \cot\bigg(\frac{\pi(j+\usa)}M\bigg) \right|=M\,\Sigma_{M,a}.
\end{equation*}
This and 
\begin{equation*}
  \sin^2(M \theta_a)=\sin^2 (\pi \sigma_a)= \sin^2 (\pi s_a)
\end{equation*}
yield (\ref{eq:ErrAvg}) as claimed. \qed
\end{proof}

%\vspace{12pt}

\noindent From Lemma
\ref{lem:ErrAvg} we see that the sum $\Sigma_{M,a}$ is the most
important part of the local average error $\ea_1(M,f)$.  We now
estimate $\Sigma_{M,a}$.

\begin{lemma}
\label{lem:ErrAvgLemma1}

Assume that $\sigma_a \notin \Z$ and $M\ge 3$. Then 
\begin{multline}
  \label{eq:ErrAvgLemma1}
  \Bigg|\, \Sigma_{M,a} - \frac1M \cot\bigg(\frac{\pi
    s_a}{M}\bigg) - \frac1M \bigg|\cot\bigg(\frac{\pi(M-1+
    s_a)}{M}\bigg)\bigg| \\  - \frac1\pi\int_{\pi (1+s_a
   )/M}^{\pi(M-1+s_a)/M} |\cot x| \, dx \; \Bigg|  \le
  \frac{1}{\pi M}\int_{\pi (1+s_a)/M}^{\pi(M-1+s_a)/M}
  \frac1{\sin^2x} \, dx.
\end{multline}
\end{lemma}

\begin{proof}
  This can be shown by applying the error formula for rectangle
  quadratures (\ref{rectangle}). 
  Note that $\pi\Sigma_{M,a} - \frac{\pi}M \cot(\pi s_a/M) -
  \frac{\pi}M |\cot(\pi(M-1+ s_a)/M)|$ is the rectangle quadrature for
  the integral $\int_{\pi (1+s_a)/M}^{\pi(M-1+s_a)/M} |\cot x| \, dx$
  with $k=M-2\ge1$. We then obtain 
  (\ref{eq:ErrAvgLemma1}) by using~(\ref{rectangle}). \qed
\end{proof}

%\vspace{12pt}

We now present the final estimate on the local average error 
$\ea_1(f,M)$.
\begin{theorem}
\label{thm:ErrAvgLoc}
Assume that $f \in \B_N$ and  $a=a_f$. For $M \ge 3$, the 
average error of the \textbf{QS} algorithm for the function $f$ satisfies
\begin{equation}
  \label{eq:ErrAvgLoc}
  \left| \ea_1(f,M) - \frac{2 \sin^2(\pi s_a) \sin(2 \theta_a)}\pi\;
  \frac{\ln M}M  \right| \le \frac{3\pi+2+\ln(\pi^2)}{M\pi}
  \;\sin(\pi s_a).   
\end{equation}
\end{theorem}

\begin{proof}
  For $\sigma_a \in \Z$ we have $s_a=0$ and (\ref{eq:ErrAvgLoc}) holds
  since $\ea_1(f,M)=0$ by~\cite{KW}.  Assume that $\sigma_a \notin
  \Z$. From Lemmas \ref{lem:ErrAvg} and \ref{lem:ErrAvgLemma1} we have
\begin{multline*}
  \bigg| \ea_1(f,M) - \frac{\sin^2(\pi s_a) \sin(2\theta_a)}{\pi
    M} \int_{\pi (1+s_a
    )/M}^{\pi(M-1+s_a)/M} |\cot x| \, dx \bigg| \\
  \le  \frac{\sin^2(\pi s_a)}M \Bigg[ 
  \frac{\sin(2 \theta_a)}M \Bigg(
  \cot\bigg(\frac{\pi s_a}{M}\bigg) 
  + \bigg|\cot\bigg(\frac{\pi(M-1+ s_a)}{M}\bigg)\bigg| \\
  + \frac1\pi\int_{\pi (1+s_a)/M}^{\pi(M-1+s_a)/M} \frac1{\sin^2x} \,
    dx \Bigg) 
+ |\cos(2 \theta_a)| \Bigg].
\end{multline*}
Observe that 
\begin{align*}
  \int_{\pi (1+s_a
    )/M}^{\pi(M-1+s_a)/M} |\cot x| \, dx &= \ln\Bigg( \sin^{-1}
    \bigg(\frac{\pi(1+s_a)}M\bigg) \sin^{-1}
    \bigg(\frac{\pi(1-s_a)}M\bigg)\Bigg),\\
    \bigg|\cot\bigg(\frac{\pi(M-1+ s_a)}{M}\bigg)\bigg| &=
\cot\bigg(\frac{\pi(1- s_a)}{M}\bigg) \le
    \cot\bigg(\frac{\pi s_a}{M}\bigg), \\
  \int_{\pi (1+s_a)/M}^{\pi(M-1+s_a)/M}  \frac1{\sin^2x} \, dx  &=
    \cot\bigg(\frac{\pi(1- s_a)}{M}\bigg) +  \cot\bigg(\frac{\pi(1+
    s_a)}{M}\bigg) \\  &\le 2 \cot\bigg(\frac{\pi s_a}{M}\bigg).
\end{align*}
The four  formulas above yield 
\begin{multline*}
\Bigg| \ea_1(f,M) - \frac{\sin^2(\pi s_a) \sin(2\theta_a)}{\pi M} \\
\times \ln\Bigg( \sin^{-1} \bigg(\frac{\pi(1+s_a)}M\bigg) \sin^{-1}
\bigg(\frac{\pi(1-s_a)}M\bigg)\Bigg) \Bigg| \\
\le \frac{\sin^2 (\pi s_a)}M \bigg( \frac{(2+2/\pi)\sin(2\theta_a)}M
\cot \bigg(\frac{\pi s_a}M \bigg) + |\cos(2\theta_a)| \bigg).
\end{multline*}
Observe that $\sin(\pi s_a)/[M\sin(\pi s_a/M)] \le 1$ since $s_a
\in (0, \frac12]$. This and the obvious estimates of sine and cosine
yield
\begin{multline}
\label{eq:aux_estim3}
\Bigg| \ea_1(f,M) - \frac{\sin^2(\pi s_a) \sin(2\theta_a)}{\pi M}\\
\times \ln\Bigg( \sin^{-1} \bigg(\frac{\pi(1+s_a)}M\bigg) \sin^{-1}
\bigg(\frac{\pi(1-s_a)}M\bigg)\Bigg) \Bigg| \\
\le \bigg(3+\frac2\pi\bigg) \frac{\sin(\pi s_a)}{M}.
\end{multline}
Consider now the left hand side of (\ref{eq:aux_estim3}). Remembering
that $M \ge 3$, and since $2x/\pi \le \sin x \le x$ for $x \in [0,
\tfrac{\pi}2]$, we get
\begin{equation}
  \label{eq:aux_estim1}
  \Bigg|\ln\Bigg( \sin^{-1}
    \bigg(\frac{\pi(1+s_a)}M\bigg) \sin^{-1}
    \bigg(\frac{\pi(1-s_a)}M\bigg)\Bigg) - 2\ln M \Bigg| \le
    \ln(\pi^2).
\end{equation} 
Thus by (\ref{eq:aux_estim3}) and (\ref{eq:aux_estim1}) we get the
final estimate (\ref{eq:ErrAvgLoc}). \qed
\end{proof}

\subsection{Worst-Average Error}
\label{sec:WAvgErr}

From Corollary \ref{thm:ErrAvgLoc_q>1} and Theorem \ref{thm:ErrAvgLoc}
we get sharp estimates on the worst-average error of the \textbf{QS}
algorithm.

\begin{theorem}
\label{thm:ErrWorAvg}
  Let $M\ge3$. Then the worst-average error of the \textbf{QS}
  algorithm satisfies the following bounds.
  \begin{itemize}

  \item For $q\in(1,\infty)$,
    \begin{equation}
      \label{eq:ErrWorAvgUpper_q}
      \ewa_q(M) \le \frac1{M^{1/q}} \bigg( \frac1{\pi} \int_0^\pi
      \sin^{q-2}(x) \, 
      dx\bigg)^{1/q}\big(1+o(1)\big). 
    \end{equation}
The last estimate is sharp, i.e.,  
\begin{equation}
  \label{eq:ErrWorAvgAsym_q}
\ewa_q(M) = \Theta\bigg(\frac1{M^{1/q}}\bigg).
\end{equation}
In particular, for $M-2$ divisible by $4$ we have
\begin{equation}
\label{eq:ErrWorAvgLower_q}
\ewa_q(M) \ge \frac1{M^{1/q}} \bigg( \frac1{\pi} \int_0^\pi
      \sin^{q-2}(x)\,|\cos(x)|^q \, 
      dx\bigg)^{1/q}\big(1+o(1)\big), 
\end{equation}
and the ratio of the integrals in (\ref{eq:ErrWorAvgUpper_q}) and
(\ref{eq:ErrWorAvgLower_q}) are approximately $1$ for $q$ close to $1$.

  \item For $q=1$,
  \begin{equation}
\label{eq:ErrWorAvgUpper}
    \ewa_1(M) \le \frac2\pi \, \frac{\ln M}M + 
        \frac{3\pi+ 2+\ln(\pi^2)}{M\pi}.
  \end{equation}
This estimate is sharp, i.e.,
\begin{equation}
  \label{eq:ErrWorAvgAsym}
 \ewa_1(M)=\Theta(M^{-1}\ln M).
\end{equation}
In particular, for $M-2$ divisible by $4$ we have 
\begin{equation*}
%\label{eq:ErrWorAvgLower}
\ewa_1(M) \ge \frac2\pi \frac{\ln M}M - 
     \frac{3\pi+ 2+\ln(\pi^2)}{M \pi}.
\end{equation*}
\end{itemize}
\end{theorem}
\begin{proof}
  Consider first the case $q \in (1,\infty)$.  By
  Corollary~\ref{cor:ErrAvgLoc_q>1} we have for all $f \in \B_N$,
  \begin{equation*}
    \ea_q(f,M) \,\le\, \frac1{M^{1/q}} \bigg(\frac1{\pi} \int_0^\pi
    \sin^{q-2}(x)\,dx\bigg)^{1/q}\big(1+o(1)\big),
  \end{equation*}
  where $o(1)$ is independent of $f$. This yields (\ref{eq:ErrWorAvgUpper_q}).

The estimate (\ref{eq:ErrWorAvgUpper_q}) is sharp since we can take a
Boolean function $f$ such that $s_{a_f} \approx \tfrac12$. Then
(\ref{eq:ErrAvgLoc_q>1}) yields (\ref{eq:ErrWorAvgAsym_q}). In particular,
for $M=4k+2$ and $a_f=1/2$ we have $\theta_{a_f} =\pi/4$, 
$\sigma_{a_f}=M/4=k+1/2$ and $s_a=\tfrac12$. Therefore 
\begin{equation*}
    \ea_q(f,M)=\frac1{M^{1/q}} \bigg( \frac1{\pi} \int_0^\pi
      \sin^{q-2}(x)\,|\cos(x)|^q \, 
      dx\bigg)^{1/q}\big(1+o(1)\big)
\end{equation*}
which proves~(\ref{eq:ErrWorAvgLower_q}). For $q$ close to $1$, the
value of $\int_0^{\pi}\sin^{q-2}(x)dx$ is mostly due to the integrand
values close to $0$ and $\pi$. Since $|\cos(x)|^q$ is then approximately
equal to one,  the ratio of the upper and lower bound integrals is
about $1$.

For $q=1$ the estimate (\ref{eq:ErrWorAvgUpper}) follows directly from
Theorem \ref{thm:ErrAvgLoc}. To prove~(\ref{eq:ErrWorAvgAsym}) it is
enough to choose a Boolean $f$ for which  the numbers
$$
\sin^2(\pi s_{a_f})\sin(2\theta_{a_f})\,=\,
\sin^2(M\theta_{a_f})\sin(2\theta_{a_f})
$$
are uniformly (in $M$) separated from $0$, see Theorem
\ref{thm:ErrAvgLoc}.  More precisely, since $a_f$ can take any value
$k/N$ for $k=0,1,\dots,N$, we take a Boolean function $f$ such that
$|a_f-\sin^2(\pi/4+\pi/(5M))|\le 1/(2N)$.  For sufficiently large $N$, we have
$\theta_{a_f}\approx \tfrac14\pi+\tfrac1{5M}\pi$.  For large
$M=4k+\beta$ with $\beta\in\{0,1,2,3\}$, we then have
$$
\sin^2(M\theta_{a_f})\sin(2\theta_{a_f})\,\approx\,
\sin^2\left(\frac{4+5\beta}{20}\pi\right)\,\sin\left(\frac12\pi +
\frac1{2.5M}\pi\right)\,>\,c\,>0,
$$
for some $c$ independent of $M$.

In particular, for $M-2$ divisible by $4$ we take $N>M$ and a Boolean
function $f\in\B_N$ with $a_f=1/2$. Then  
$$
s_a=\tfrac12\quad\mbox{and}\quad \sin^2 (\pi s_a) \sin(2\theta_a)\,=\,1,
$$
which leads the last estimate of Theorem \ref{thm:ErrWorAvg}.  \qed
\end{proof}

\subsection{Quantum Summation Algorithm
  with Repetitions} 
\label{sec:repete}

The success probability of the \textbf{QS} algorithm is increased by
repeating it several times and taking the median of the outputs as the
final output, see e.g.,~\cite{H1}.  We show in this section that this
procedure also leads to an improvement of the worst-average error
estimate.

We perform $2n+1$ repetitions of the \textbf{QS} algorithm for some $n
\in \{0,1,\ldots\}$. We obtain $\sin^2(\pi j_1/M),\sin^2(\pi
j_2/M), \ldots,\sin^2(\pi j_{2n+1}/M)$ and let $\bar a_{n,f}$ be the
median of the obtained outputs, i.e., the $(n+1)$st number in the ordered
sequence. Let $\rho_{n,f} (\alpha)$, $\alpha \in \mathcal{A}_M$, be
the probability that the median $\bar a_{n,f}$ is equal to~$\alpha$.
This probability depends on the distribution function $F_f$ of the
original outputs from $\mathcal{A}_M=\{\sin^2(\pi j/M):
j=0,1,\ldots,M-1\}$, which is defined as
\begin{equation*}
  F_f(\alpha)=
\begin{cases}
\sum_{\alpha' \in \mathcal{A}_M, \alpha' < \alpha }  \rho_f(\alpha')
 &\text{for $\alpha > 0$},\\
0 &\text{for $\alpha = 0$}.
\end{cases}
\end{equation*}
It is known, see \cite{STAT} p. 410, that the distribution of the
median $\bar a_{n,f}$ is of the form
\begin{equation}
  \label{eq:median-distr}
   \rho_{n,f}(\alpha) = (2n+1) \binom{2n}n
  \int_{F_f(\alpha)}^{F_f(\alpha) + \rho_f(\alpha)}  t^n(1-t)^n \,
  dt,\quad \forall \alpha \in \mathcal{A}_M. 
\end{equation}

%\vspace{12pt}

We are now ready to estimate the worst-average error of the \textbf{QS}
algorithm with $2n+1$ repetitions 
\begin{equation*}
  \label{eq:local-avg-err-rep}
 \ewa_{q,n}(M) = \sup_{N>M}\, \max_{f \in \B_N} 
\bigg(\sum_{\alpha \in
   \mathcal{A}_M}  \rho_{n,f}(\alpha) |a_f-\alpha|^q
   \bigg)^{1/q},\qquad q \in [1,\infty).   
\end{equation*}

We estimate $\ewa_{q,n}(M)$ by using Theorem 12 of \cite{BHMT} which
states that the \textbf{QS} algorithm with $M$ queries computes
$\bar a_f$ such that
$$
|a_f-\bar a_f|\,\ge\, c_1\frac{k}{M} \quad \mbox{with probability
  at most} \quad \frac{c_2}k
$$
for any positive integer $k$, Here $c_1$ and $c_2$ are absolute
constants and $f$ is any Boolean function from $\B_N$.
If
$$
|a_f-\bar a_{n,f}|\,\ge \, c_1\, \frac{k}{M},
$$
then for at least $n$ outcomes $\bar a_f(j_1),\ldots, \bar a_f(j_n)$
we must have
$$
|a_f-\bar a_f(j_l)|\, \ge\, c_1\frac{k}{M} \quad \mbox{for}\ l=1,\dots, n.
$$
But the probability that this occurs is bounded by 
$$
{2n+1 \choose n}  \left(\frac{c_2}{k}\right)^n.
$$
It follows then that (with $c$ which may depend on n)
$$
\mbox{Prob}\,\{\,|a_f-\bar a_{n, f}|\,\ge\, c_1k/M\,\}\, \le\, c\,k^{-n}.
$$
We now use the standard summation by parts. Define
$$
p_k\,=\,\mbox{Prob}\,\{\,c_1(k-1)/M\,\le\,|a_f-\bar a_{n,f}|\,<\, c_1k/M\}.
$$
Then, by the above estimate we get for an arbitrary integer $l$, 
$$
\sum_{k>l}p_k\,\le\, c\,l^{-n}.
$$
Therefore
\begin{eqnarray*}
\ewa_{q,n}(M)^q&\,\le\,&\sum_{k=1}^\infty p_k (c_1k/M)^q\,=\,
(c_1/M)^q\,\sum_{k=1}^\infty p_k \,\sum_{l=1}^k(l^q-(l-1)^q)\\
&\,=\,&(c_1/M)^q\,\sum_{l=1}^\infty(l^q-(l-1)^q)\,\sum_{k=l}^\infty p_k\\
&\,\le\,& cM^{-q}\sum_{l=1}^\infty l^{q-1-n}\,\le\ c\,M^{-q}
\end{eqnarray*}
for $n>q$ and with the number $c=c_{q,n}$ depending only on $q$ and
$n$.  In fact, taking $n=\lceil q\rceil+1$ it is easy to check that
$c_{q,n}$ is a single exponential function of~$q$. Hence, by taking
the $q$th root we have 
\begin{equation*}
  \ewa_{q,n}(M)\le c_{q,n}^{1/q} M^{-1}
\end{equation*}
with $c_{q,n}^{1/q}$ of order $1$. Therefore we have proven the
following theorem.

\begin{theorem}\label{threp}
The worst-average error of the median of $2(\lceil q\rceil +1)+1$ repetitions 
of the \textbf{QS} algorithm with $M$ quantum queries satisfies
$$
\ewa_{q,\lceil q\rceil+1}(M)\,=\,O(M^{-1})
$$
with an absolute constant in the big $O$ notation 
independent of $q$ and $M$. \qed
\end{theorem}

The essence of Theorem \ref{threp} is that the number of repetitions
of the \textbf{QS} algorithm is {\it independent} of $M$ and depends
only linearly on $q$.  Still, it allows to essentially improve the
worst-average error of the \textbf{QS} algorithm. As we already
mentioned in the introduction, the bound of order $M^{-1}$ is a lower
bound on the worst-average error of any quantum algorithm.  Hence, the
\textbf{QS} algorithm with repetitions enjoys optimality also in the
worst-average setting.
 
\vspace{24pt}

\begin{center}
\textbf{ACKNOWLEDGMENTS}
\end{center}
We wish to thank P. H\o{}yer for suggesting to study repetitions of
the \textbf{QS} algorithm for $q=1$. We are also grateful to
J.~Creutzig, E.~Novak, A.~Papageorgiou, J.~F.~Traub and
A.~G.~Werschulz for valuable comments.

\end{document}